\documentclass[10pt]{emulateapj}
\usepackage{apjfonts}
\usepackage{psfig}
\usepackage{epsfig}
\usepackage{hyperref}

\hypersetup{
    pdfauthor   = {Nicole P. Vogt},
    pdftitle    = {A New Resource for College Distance Education Astronomy Laboratory Exercises},
    pdfsubject  = {Astronomy},
    pdfkeywords = {astronomy education research; astronomy education resources; college-level astronomy; distance education; laboratory exercises; STEM coursework},
    colorlinks  = true,
    linkcolor   = blue,
    filecolor   = magenta,      
    urlcolor    = cyan
}

\lefthead{Vogt, Cook, \& Muise}
\righthead{A New Resource for College Distance Education Astronomy Laboratory Exercises}

\slugcomment{Accepted by American Journal of Distance Education 2012 September 17}

\begin{document}

\title{A New Resource for College Distance Education Astronomy Laboratory Exercises}

\author{Nicole P. Vogt$^1$, Stephen P. Cook$^1$, and Amy Smith Muise$^1$}

\affil{New Mexico State University, Department of Astronomy, \\
       P.O. Box 30001, Dept 4500, Las Cruces, NM 88003}
\email{nicole@nmsu.edu}

\begin{abstract}

This article introduces a set of distance education astronomy laboratory exercises for use by college students and instructors and discuss first usage results.  This General Astronomy Education Source (GEAS) exercise set contains eight two-week projects designed to guide students through both core content and mathematical applications of general astronomy material.  Projects are divided between hands-on activities and computer-aided analyses of modern astronomical data.  The suite of online resources includes student and instructor guides, laboratory report templates, learning objectives, video tutorials, plotting tools, and web-based applications that allow students to analyze both images and spectra of astronomical objects.  A pilot usage study indicates that distance learners using these materials perform as well or better than a comparison cohort of on-campus students.  We are actively seeking collaborators to use these resources in astronomy courses and other educational venues.

\end{abstract}


\section{Introduction}

College distance education is experiencing substantial growth: the number of participating undergraduates rose by 21\% in 2009-2010 alone, while in the same period the overall undergraduate population saw only a 2\% increase (\hyperlink{Allen2010}{Allen \& Seaman 2010}).  However, the physical sciences lag fields such as computer science and business by a factor of two in this arena (\hyperlink{Radford2012}{Radford 2012}). Particular challenges arise for distance educators in the physical sciences, perhaps most significant the difficulty of conducting laboratory exercises remotely, a problem highlighted in a recent special topics issue of this journal (\hyperlink{Kennepohl2009}{Kennepohl 2009}) devoted to teaching science. Various approaches (\hyperlink{Ma2006}{Ma \& Nickerson 2006}) -- including mobile and regional laboratories (\hyperlink{VanMil2010}{Van Mil et al. 2010}), home experiment kits (pioneered by the Open University in the 1970s, as described by (\hyperlink{Perry1976}{Perry [1976]}), remote laboratories (cf. (\hyperlink{Gravier2008}{Gravier 2008}; (\hyperlink{Eckert2009}{Eckert, Gr\"ober, \& Jodl 2009}), and computer simulations (\hyperlink{Kennepohl2001}{Kennepohl 2001}) -- have been pursued in the effort to provide students a meaningful laboratory experience at a distance.  

Among the physical sciences, the field of astronomy offers several advantages in the distance education realm. Astronomy research is by nature conducted remotely, requiring neither detailed treatments of physical samples (as in chemistry) nor field work (as in geology).  Astronomers use imaging cameras and spectroscopes to observe planets, stars, and galaxies, tracing photons across the cosmos but rarely establishing physical contact with these distant objects of study.  This trend is accentuated in modern astronomy, with satellite and spacecraft-based projects such as the National Aeronautics and Space Administration (NASA) Hubble Space Telescope and Mars Global Surveyor, and collaborative projects like the ground-based Sloan Digital Sky Survey (SDSS) \hyperlink{Aihira2011}{(Aihara et al. 2011)}, which allow astronomers spread across the globe to analyze observational data archived and shared on a nightly basis.  

Additionally, the astronomy education community has long maintained a rich presence online. Reservoirs of astronomy materials for online usage are well organized and often freely available.  These include the work of the Contemporary Laboratory Experiences In Astronomy (CLEA) (\hyperlink{Marschall2000}{Marschall, Snyder, \& Cooper 2000}), an extensive archive of laboratory exercises that can be downloaded and executed on Windows PCs; computer simulations of astronomical behavior created by the Nebraska Astronomy Applet Project (\hyperlink{Lee2012}{Lee 2012}) and the Virtual Laboratory at the University of Oregon (\hyperlink{Bothun2012}{Bothun 2012}); and the SDSS set of advanced projects (\hyperlink{Raddick2002}{Raddick 2002}) outlining topics in stellar and extragalactic astronomy.  

Offerings of true distance education laboratory experiences in astronomy, however, are rare. We report here on a new resource for distance education astronomy laboratory exercises, providing both hands-on and platform-independent computer-based experiments that illustrate a range of general scientific skills and techniques to develop scientific literacy (\hyperlink{NRC1996}{National Research Council 1996}) and also showcase the rich data available in modern astronomical research.  A suite of supporting materials is designed to enhance accessibility for instructors with limited local resources.  We also discuss the results of a pilot study indicating that distance learning students using these laboratory exercises did as well as or better than on-campus peers.  

\section{Description of GEAS Laboratory Projects}

\begin{table*} [htbp]
  \caption{List of Laboratory Exercises}
  \begin{center}
  \begin{tabular} {l c c} 
  \hline
  \hline
  Project Title & M$^1$ & Lecture Modules \\
  \hline 
  1. Fundamentals of Measurement and Error Analysis & HA & 2, 9, 11 \\ 
  2. Observing the Sky & HA & 4-5, 7-8 \\
  3. Cratering and the Lunar Surface & HA & 6, 11-12 \\
  4. Cratering and the Martian Surface & OB & 11-12 \\
  5. Parallax Measurements and Determining Distances & HA & 7-8, 10 \\
  6. The Hertzsprung-Russell Diagram and Stellar Evolution & OB & 17-18, 20-21 \\
  7. Hubble's Law and the Cosmic Distance Scale & OB & 19, 24-26 \\
  8. Galaxy Properties & OB & 16-17, 19, 24-26 \\
      ~~~~~Stellar and Quasar Spectral Analysis Tools & OB & 14-17, 20 \\
 \hline 
\multicolumn{3}{l}{$^1$ Project mode (HA=primarily hands-on, OB=primarily observational data-driven)} \\
  \end{tabular}
  \end{center}
  \label{tab01}
\end{table*} 

The General Education Astronomy Source (\href{http://astronomy.nmsu.edu/geas/}{GEAS}) project at New Mexico State University has created a sequence of eight astronomy laboratory projects, each designed to take place over a two-week period.  The first project illustrates how to design, conduct, and analyze an experiment, and introduces statistical and plotting tools that can be used to analyze a wide range of experimental data.  We are aware that 40\% of the students in introductory science courses intend to become teachers (\hyperlink{Lawrenz2005}{Lawrenz, Huffman, \& Appeldoorn 2005}), so our intent is both to teach our current students how to explore and solve problems scientifically and to leverage our efforts by preparing future teachers to disseminate this knowledge at the K-12 level.  Once students are grounded in experimental design and basic statistics, the remaining projects are split fairly evenly between solar system exercises and stellar and extragalactic exercises.  This provides a wide range of options from which instructors may select, for example, six of eight projects for a given semester in a general astronomy course.  

Four of the projects are primarily hands-on, focused on experiments that students conduct and analyze for themselves, and four are supported by archives of astronomical imaging and spectral observational data.  All project communication and materials distribution can be conducted via the Internet; we do not send kits through the mail, and student reports and instructor feedback are shared electronically.  We are also conscious of the benefit to extending the student learning experience beyond pure hypermedia (\hyperlink{Dillon1998}{Dillon \& Gabbard 1998}, \hyperlink{Dillon2005}{Dillon \& Jobst 2005}) and employ in-person experiments and peer collaborations and discussion sessions to promote social interactions.  

Our hands-on experiments have been designed so that they can be completed with common household materials such as cardboard, pins, rulers, and marbles. Students quickly master the necessary computer skills -- simple, custom-designed statistical, plotting, and data analysis tools ensure that students spend their time on astronomy and scientific experimentation rather than struggling to install software packages or working through compilation procedures.  Our resources are provided through platform-independent methods, so that students can work with a computer and operating system of their choice.  They do not need to install any client-side software.  This enables students to complete all coursework using public computers such as those found in libraries and universities, or from military bases and other workplaces.

Table 1 lists the eight laboratory projects.  The column labeled ``lecture modules'' connects the material covered in each project to a series of 26 lecture modules available to students through a library of lecture slides, audio recordings, and 12,000+ self-review questions suitable for a parallel non-laboratory course component (\hyperlink{Vogt2015}{Vogt \& Muise 2015}).  These resources are not required to complete the laboratory projects, which are self-contained.  However, when instructors elect to use the laboratory projects their students also receive access to these companion resources.  Note that lecture module 11 contains a review of several statistical topics introduced in project one, including histograms, mean values and standard deviations, and linear fits. 

\vspace{0.40truein}
\section{Project Components}

\begin{figure*} [htbp]
  \begin{center}\epsfig{file=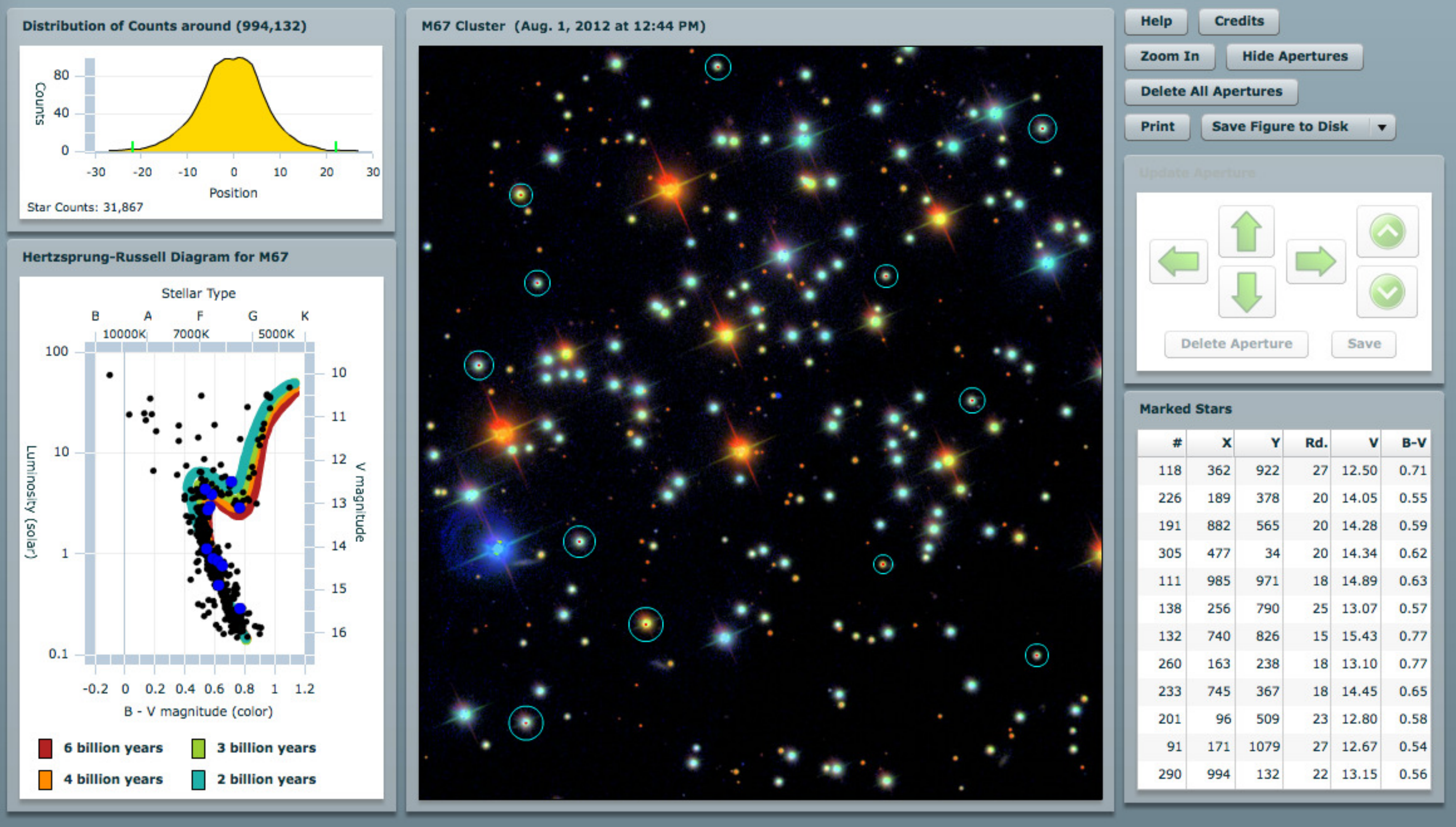, width=5.0in}\end{center}
  \caption[Figure 1] 
{Figure 1 shows our Hertzsprung-Russell (H-R) diagram and stellar aperture fitting web-application.  The radial profile in the upper left-hand corner shows the distribution of light around the star currently being fit on the cluster image. Students can vary the aperture size and position with the right-hand side control panel, using aperture controls that match those used in project four to fit craters on the Martian surface.   The H-R diagram in the lower left-hand corner shows the distribution of stellar luminosities and colors for stars selected and fit by students.  The four shaded tracks illustrate the paths followed by stars evolving off of the Main Sequence within clusters of various ages.  After students fit twelve stars, the rest of the cluster members are added to the H-R Diagram to give students a large sample when selecting an age track.}
  \label{fig01}
\end{figure*}

\subsection{Manual Chapters}

Each laboratory project contains a suite of resources for students and instructors.  The heart of each exercise is a 30-page manual chapter that introduces key concepts, describes the experiments, and defines the project questions and tasks to be completed.  Each project is designed to be a self-contained exercise; the necessary astronomical concepts are fully explained within the chapter.  Instructors using the default project ordering will find that early chapters build knowledge used in later chapters: The statistical and plotting tools for the entire sequence are explained in detail in the first chapter, while the Martian cratering studies in chapter 4 build on the lunar cratering results of chapter 3, and chapters 7 and 8 both utilize basic information regarding galaxies. However, beyond beginning with project one, the rest of the order may be freely changed, or only a subset of projects used, as desired. 

Each project centers on activities that involve hands-on experimentation or studies of astronomical images and spectra.  A series of questions placed throughout the chapter help students to understand why they complete various steps, and a set of final questions draws together concepts from throughout each project.  Each chapter ends with a summary in which students discuss the most important concepts of the project and a short extra credit activity that extends the project scope beyond the required material.  

A complete manual is available containing all eight projects, a glossary of scientific keywords used throughout, and a list of supplies for conducting all experiments.  The complete manual runs to 270 pages; shorter versions containing a subset of the eight projects are available on request.  

\subsection{Instructor Guides}

Each project contains an instructor's guide with an overview, learning objectives, definitions of relevant scientific keywords, data references and additional astronomical resources, and advice for teaching each project (gathered from the developers and pilot instructors).  Learning objectives range from project-specific astronomical concepts (such as understanding the concept of parallax, and explaining how it can be used to determine the distances to solar system and galactic objects, or relating the observed altitude of the North Star to an observer's latitude) to general scientific and technical skills (such as reading values from a logarithmic plot, and understanding how it differs from a linear plot, or understanding the effects of both measurement error and systematic error on data).  There is an average of eight for each project, split 40/60 between general science objectives and those specific to astronomical and astrophysical topics.

\subsection{Video Tutorials}

A 10- to 30-minute video tutorial illustrates the components of each project that have proven the most difficult to convey to distance learning students.  They show students and actors constructing pieces of equipment and conducting sky observations and other experiments, emphasizing proper technique and safety.  We also cover key astronomical concepts, showcase student dialogs and skits to illustrate the thinking process for understanding and testing key concepts, and provide animated simulations to help students to visualize actions such as the orbit of the Moon around the Earth, the change in observed parallax angle for nearby and distant objects, and the appearance of a galaxy from different viewing angles.  All videos are provided at no charge via a professional streaming web site (http://vimeo.com) for maximum accessibility at low bandwidths.  

\subsection{Laboratory Reports}

A template laboratory report has been created for each project, within the Google Documents template library.  Each template contains concise forms of the lab questions, drawn from the project chapter.  Spaces are defined appropriately for short and long answers, so that when the report is completed student responses appear in red and can be easily found between the prompts. Google allows students to write text, create tables, and upload figures into the documents.  The user interface is analogous to that for MSWord or Pages on a Mac and feels familiar to most students.  The numerous figures that students create with our software are optimized for small size so that they can be easily stored on disk, transferred via email if desired, and included in Google Documents without difficulty.

Students copy each template to their own Google Documents account and then share them with their instructors.  This allows instructors to see the laboratory reports over the entire two-week project interval, not just at the end.  Students are strongly encouraged to enter all experimental data into tables as the data are collected, to allow instructors to provide guidance and identify any significant errors that may appear during first attempts.  A built-in chat function allows students and instructors to trade comments back and forth in real time in the margin of the laboratory report, and asynchronous comments (with threaded replies) can be entered into the reports at specific locations where students have questions or difficulties have occurred.  

The goal is to create an interactive process of exploration and advancement between students and instructors as the data are taken, analyzed, and reported for each project.  This is particularly critical for distance learning cohorts, which lack access to the easy back-and-forth discussions which characterize the best of in-class peer groups (\hyperlink{Astin1993}{Astin 1993}; \hyperlink{Mosse2010}{Mosse \& Wright 2010}) and the instantaneous instructor guidance that the on-campus forum can deliver so well.  

\begin{figure*} [htbp]
  \begin{center}\epsfig{file=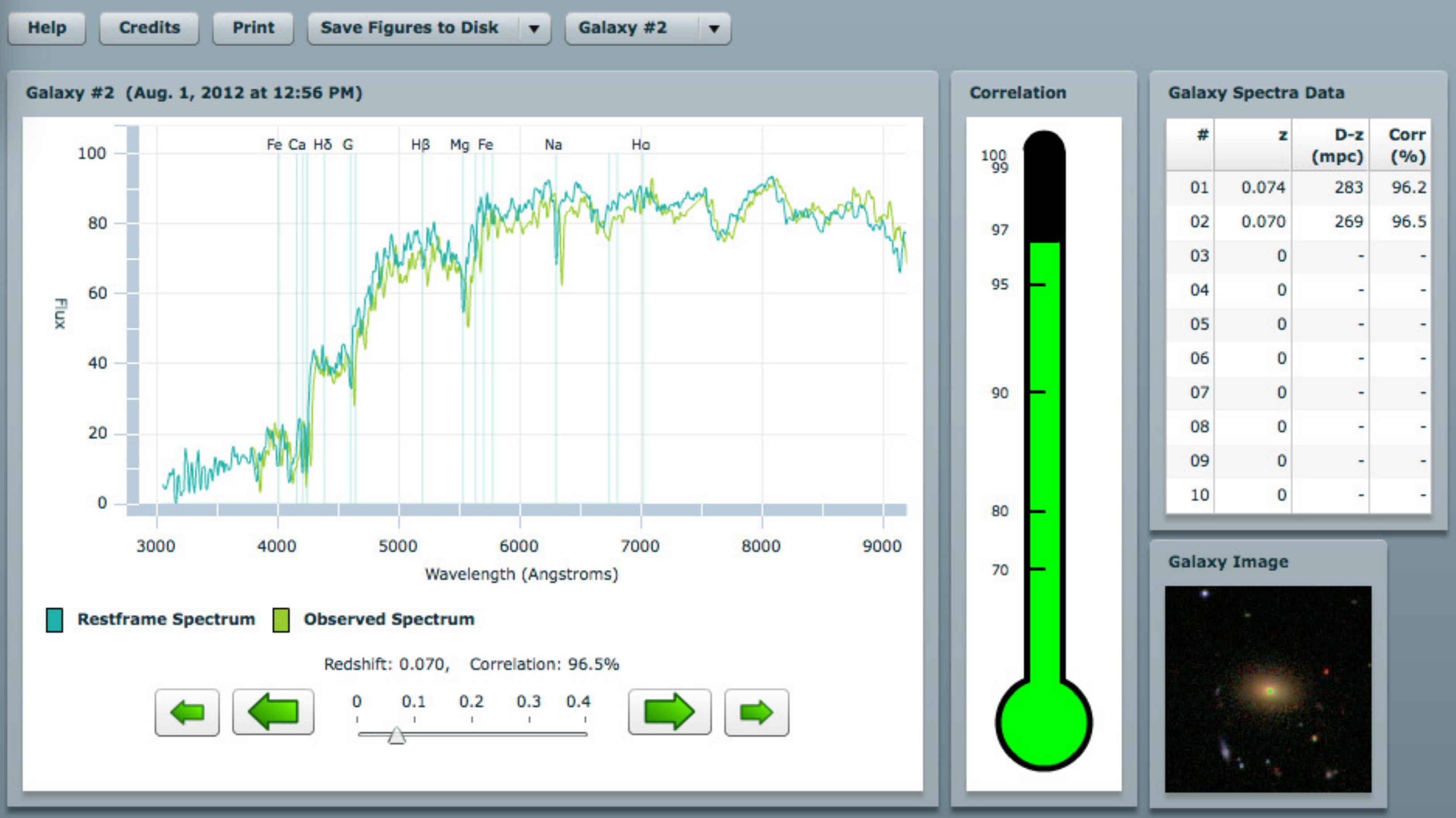, width=5.0in}\end{center}
  \caption[Figure 2] 
{Figure 2 shows our galaxy spectrum analysis tool.  Students determine galaxy redshifts by shifting the galaxy spectrum left and right in wavelength so as to match up absorption and emission features, as well as the general shape of the continuum, to a rest-frame spectrum of an equivalent galaxy.  The correlation coefficient is reported numerically and also shown graphically on a symbolic thermometer -- the better the match between the two spectra, the higher the ``liquid'' level in the thermometer. }
  \label{fig02}
\end{figure*}

\subsection{Additional Project Resources}

Each lab project has a dedicated web site with links to all relevant materials. Additional resources include data tables for particular experiments, web-applications with pre-loaded astronomical images and spectra, and plotting tools.  We also provide short guides focused on mathematical tools (correlation coefficients, common xy relationships, logarithmic plots, mean values, measurement techniques, percent differences, and slopes and y-intercepts), astronomical tools (angular extents, lunar phase wheels, and astronomical keywords), and computing aids (Google Documents).  

Students have access to a general plotting tool that provides options for creating histograms and linear plots of xy data with and without attached errors. Histograms are labeled with mean values and standard deviations.  Unweighted linear fits are labeled with slope and y-intercept values as well as correlation coefficients indicating the tightness of fit.  For weighted fits, error bars are attached along the y-axis and standard deviations shown for the slope and y-intercept values.  All plots are presented in PNG format (ideal for plots of discrete points, and generating files of small size) and labeled with student name and the time at which the plot was created.  In service of statistical literacy, we encourage students to utilize this tool for any purpose -- to create plots of data for astronomical classes, of course, but also on their own for other classes or personal investigations.  We also provide students with specialized plotting tools for specific experiments.

Our plotting tools have simple interfaces in which students select options from drop-down menus, enter text in boxes to define captions and labels, and enter numerical values to define data points for plotting.  The interfaces apply basic checks to the input data (e.g., have students entered pairs of numbers for an xy plot?); then server-side routines create an output PNG-format plot, with limits and point styles set automatically.  

Our fourteen platform-independent web applications are designed so that students need not install software, compile packages, or learn complicated commands in order to examine data and conduct basic analyses, allowing users with no background in data processing to immediately focus on astronomical issues.  

\begin{table*} [htbp]
  \caption{Comparison of Classroom and Remote Astronomy Courses}
  \begin{center}
  \begin{tabular} {l c c c} 
  \hline
  \hline
  Cohort & Scores$^1$ & Enrollment & Withdrawn \\
  \hline 
  Classroom & ~~~~~~$24.7 \pm 3.6$~~~~~~ & 85 & 14 \\ 
  Remote & $28.8 \pm 1.8$ & 21 & 14 \\ 
 \hline 
\multicolumn{4}{l}{$^1$ Scores (out of 30 points) on laboratory-based portion of final exam} \\
  \end{tabular}
  \end{center}
  \label{tab02}
\end{table*}

\section{Descriptions of Projects}

The first project introduces concepts and statistical and plotting tools used throughout the entire project sequence.  It emphasizes productive scientific experimentation -- how to design, conduct, and evaluate an experiment.  There are four activities: planning and conducting a short experiment with common household items, examining existing data to uncover a basic connection between seasonal changes and the height of the Sun in the sky at noon, analyzing data (including error estimates), and making appropriate conclusions based on evidence.  

Project two is focused on naked eye observations of the sky -- visualizing the positions of the Moon, Earth, and Sun throughout the monthly lunar orbit, and relating the changing appearance and position of the Moon in the sky to its phase, illumination and elongation angle.  Students construct sextants, and estimate their own latitude from observed altitudes of the transiting first quarter Moon and Polaris.  They build on general skills introduced in project one by interpreting figures, planning and conducting observations, recording data accurately, and analyzing the results. This project contains multiple activities that encourage students to develop the skill of visualization.  

The third project begins a two-project module exploring evolutionary processes on terrestrial surfaces.  It contains an experimental cratering activity, an analysis component in which students fit several physical models to their data and determine which one best reproduces the observed trend between projectile velocity and crater size, and a cratering simulation which enables them to predict the effects of various solar system projectiles impacting here on Earth.   

Project four shifts from the Moon to Mars, exploiting an extensive NASA reservoir of high-resolution Martian surface imaging.  These images cover regions dominated by volcanic activity, water-carved features, and water floods caused by volcanic activity.  The cratering record (the surface density of craters of various sizes) is used to determine surface ages and chart evolutionary histories.  

The fifth project centers on a hands-on activity where students build transits and conduct parallax measurements, measuring angular shifts in local object positions based on changes in observer vantage point and then connecting their experiment to larger scale parallax measurements conducted on semi-yearly timescales to measure analogous shifts for nearby Milky Way stars.  

Project six focuses on properties and evolution of stars.  It uses the Hertzsprung-Russell (H-R) Diagram to trace how stellar properties such as brightness and color change with age. Students construct their own H-R Diagram (see Figure 1) by fitting apertures to the light profiles of individual stars within a cluster. 

The seventh project is our first extragalactic project, highlighting the immense scale of the Universe.  It addresses the challenge of determining distances on large scales, working from standard candle techniques (comparing observed brightness and size for objects with uniform physical properties placed at varying distances) to cosmological redshifts. Students analyze galaxy images and spectra (see Figure 2), and use Hubble's Law to relate observed shifts in the wavelengths of spectral features to determine distances to nearby galaxies. 

Project eight focuses on galaxy properties and their constituent stellar populations.  We introduce the galaxy morphological sequence and augment luminosities, sizes, colors, and redshifts with modern indices based on concentration and asymmetry of light.  Students analyze images and spectra of a sample of nearby galaxies, determining intrinsic properties and estimating morphological types.  

\section{Spectral Analysis Tools}

In addition to our eight full-fledged laboratory projects, we have developed several additional tools for use in presenting and analyzing astronomical spectra.  The first tool is a web-application for analyzing quasar spectra (spectra of bright cores of distant galaxies).  It contains an archive of 1,300 moderate-resolution optical spectra from the Astronomy Research Based Science Education (\hyperlink{NOAO2009}{National Optical Astronomy Observatory 2009}) project (see Figure 3), and users can also load additional spectra of their own. Focus mode allows them to zoom in on individual line features, as shown in Figure 4.  They can remove the continuum level on each side of a line and fit it with a Gaussian profile. The line diagnostics (e.g., equivalent width comparisons) can support detailed analyses of quasar properties. 

\begin{figure*} [htbp]
  \begin{center}\epsfig{file=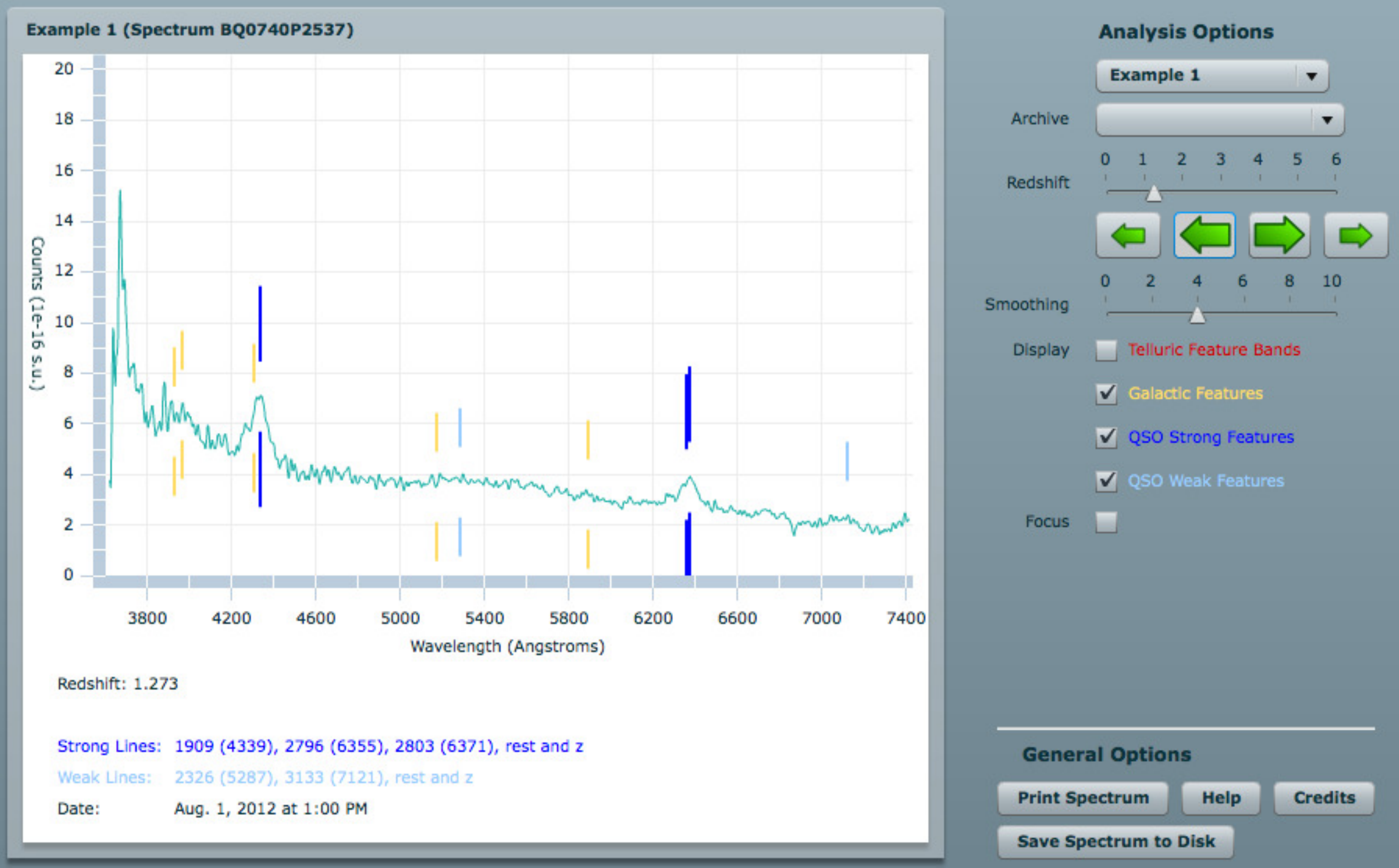, width=5.0in}\end{center}
  \caption[Figure 3] 
{Figure 3 shows the primary screen for our quasar fitting tool.  Redshift-dependent line features (due to light from the quasar) are identified with heavy bars, and Galactic features with light bars.  Students can vary the quasar redshift to line up specific spectral features with known quasar lines, while disregarding features caused by Milky Way and atmospheric factors.  They can also smooth the spectrum in wavelength, to decide whether bumps and wiggles in the spectrum are real or noise.}
  \label{fig03}
\end{figure*}

\begin{figure*} [htbp]
  \begin{center}\epsfig{file=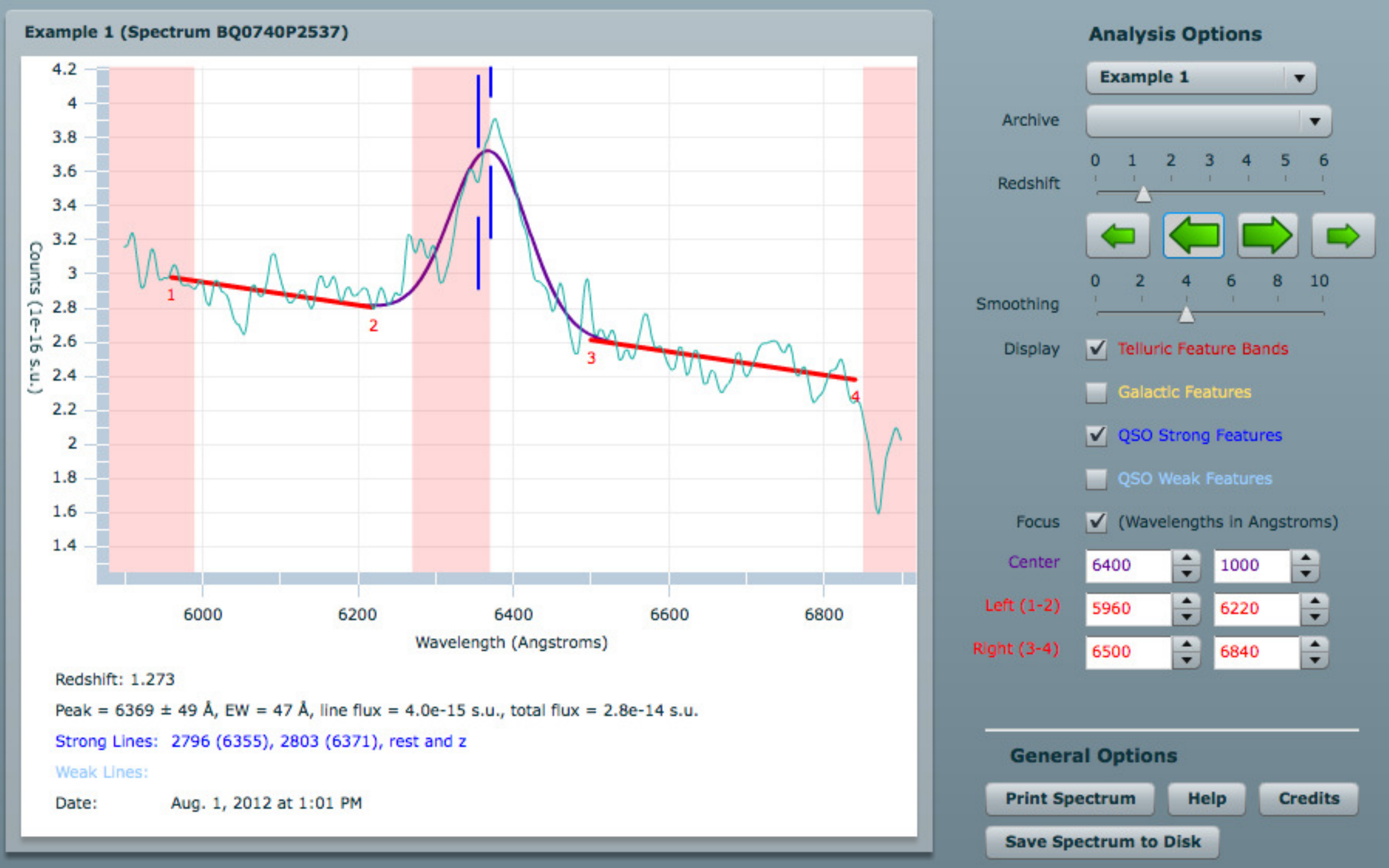, width=5.0in}\end{center}
  \caption[Figure 4] 
{Figure 4 shows the focus mode screen for our quasar fitting tool.  Users can define the level of the continuum on both sides of a line feature (interactively with the mouse or keyboard controls, or by setting wavelength limits numerically).  The tool then subtracts off the continuum level of flux and fits a Gaussian line profile to the line, returning the central wavelength and line width as well as the equivalent width and line flux.}
  \label{fig04}
\end{figure*}

A second tool supports the analysis of stellar spectra.  It echoes many of the features of the quasar line-fitting tool, allowing the user to overlay marks for elemental or molecular features as well as Telluric bands (atmospheric effects).  A temperature slider bar is used to define the blackbody temperature and peak wavelength of the star. This tool can be used for a broad range of projects involving the study of stellar spectra once an introduction has been made to the basics of stellar types and spectroscopy. 

\section{A Comparison of Laboratory Project Usage Outcomes (Remote and On-Campus Cohorts)}

We have data for two general education astronomy courses taught by the same instructor using our laboratory exercises, one in the classroom in Fall 2010 and one via distance learning in Fall 2011.  Because exams were not returned to students, we were able to use an identical 30\% of the two-hour final exam for both cohorts to cover material studied during laboratory projects.  The mean values and standard deviations for this portion of the final exam are shown in Table 2.  The distance learning cohort in fact did significantly better on this material, with a mean value more than one pooled standard deviation above that for on-campus students, and a two-sample t-test yields a less than 1\% probability that the two results were drawn from the same parent sample.  The remote laboratory projects thus appear effective (note that students worked individually in both cohorts).

We emphasize that the distance learning class was a small pilot cohort and had a large fractional withdrawal rate (four times that of the on-campus course).  Students who completed the course and those who withdrew performed equally on a pre-class survey of astronomical concepts (the results were also not significantly different than those for the on-campus cohort), and reported equal experience with computers.  Those who completed the course via distance learning were older (25-30 years old, rather than 20 or younger) and more than twice as likely to have children (as expected, cf. \hyperlink{Radford2012}{Radford 2012}), and had studied math beyond algebra (though their self-assessed math skills were not significantly higher) rather than stopping at pre-algebra or algebra.  However, half of the students who withdrew did so without completing any course work, most citing difficulties securing financial aid as their primary cause for leaving, rather than any course-related issues.  Other common reasons mentioned were lack of time to study, transfer to an on-campus course, or birth or death in the family.  While more study is needed, our initial results are promising and suggest that (1) distance learning students may be learning as much or more through the laboratory projects than on-campus students, and (2) exposure to math beyond algebra is helpful for course success.

\section{Examples of Laboratory Project Modes}

\subsection{Semester- or Term-Long Class Sequence}

The first mode applies to a semester-long class in which students complete up to eight two-week projects.  Instructors are provided with a PDF-format laboratory manual with the necessary materials.  Students receive individual accounts for plotting and statistical tools and have complete access to the suite of materials available online. This model works well for faculty without access to an existing set of laboratory activities or lacking resources to support a laboratory program in-house.  All projects can be conducted in-class, via distance learning, or through a hybrid model.  

\subsection{Single Project in Stand-Alone Mode}
The second mode applies to an instructor wishing to supplement an existing set of laboratory exercises with several from our GEAS sequence, teaching a class without a full-fledged laboratory component and seeking to conduct one or more GEAS projects as class or homework assignments, or wanting to utilize a GEAS project for outreach.  

We provide laboratory chapters for individual projects, and instructors can choose whether to request student accounts for online tools.  We have alternative versions for several projects that circumvent the need for students to use the plotting resources individually.  Note that we also provide short guides for much of the statistics material introduced in project one, useful for instructors who just wish to use one of the other projects.

\vspace{0.20truein}
\subsection{Plotting, Imaging Analysis, or Spectrum Analysis Tool Usage}
The third mode applies to individuals who wish to use our platform-independent tools to support independent astronomical data analysis projects.  The most likely tools for independent use are the spectral analysis tools discussed in \S5, or the general plotting tool discussed in \S3.5 (providing histograms, mean values and standard deviations, and weighted and un-weighted linear fits).  We also provide a third spectral analysis tool, for analyses of galaxy spectra.  We are interested in collaborations to construct astronomy educational and outreach activities that leverage these resources.  

\vspace{0.20truein}
We are currently soliciting collaborators, and welcome inquiries from instructors interested in working with our materials for distance education cohorts, in the classroom, via hybrid models, and for use in outreach programs.  Our materials are copyrighted, but are available by request to instructors at non-profit educational institutions and outreach organizations.  Individual test accounts for the plotting resources are also available by request.  

\section{Resources}

More information on the General Education Astronomy Source (GEAS) may be found online at \href{http://astronomy.nmsu.edu/geas/}{http://astronomy.nmsu.edu/geas}, or by contacting us at geas@astronomy.nmsu.edu. 


\acknowledgements

We thank our pilot program students and instructors from New Mexico State University.  This material is based in part upon work supported by the National Science Foundation (NSF) through Grant No. AST-0349155 to NPV and by NASA through Grant No. NNX09AV36G to NPV. Any opinions, findings and conclusions or recommendations expressed in this material are those of the authors and do not necessarily reflect the views of NSF or NASA.

\end{document}